\documentclass[runningheads]{llncs}
\usepackage[utf8]{inputenc}
\usepackage{todonotes}
\usepackage{booktabs}
\usepackage{caption}
\usepackage{subcaption}

\newcommand{\figref}[1]{Fig.~\ref{#1}}
\newcommand{\tabref}[1]{Tab.~\ref{#1}}

\newcommand{\omp}[0]{OpenMP\,5.0\,}

\title{
 Task inefficiency patterns for a wave equation solver
 }

\titlerunning{Task inefficiency patterns}
\author{
 Holger Schulz\inst{1}
 \and 
 Gonzalo Brito Gadeschi\inst{2}
 \and
 Oleksandr Rudyy\inst{3}
 \and
 Tobias Weinzierl\inst{1}
}
\authorrunning{H. Schulz et al.}
\institute{
    Department of Computer Science,
 Durham University, United Kingdom\\
 \email{\{holger.schulz,tobias.weinzierl\}@durham.ac.uk}
\and
NVIDIA GmbH, Munich, Germany \\\email{gonzalob@nvidia.com}
\and
High Performance Computing Center Stuttgart (HLRS), University of Stuttgart, Germany\\\email{oleksandr.rudyy@hlrs.de}}
\date{April 2021}

\usepackage{graphicx}
\usepackage{algorithm}
\usepackage{algpseudocode}

\newtheorem{observation}{Observation}
\newtheorem{feature}{Feature}

\begin{document}

\maketitle

\begin{abstract}
  The orchestration of complex algorithms demands high levels of automation to
use modern hardware efficiently. Task-based programming with \omp is a
prominent candidate to accomplish this goal. We study \omp's tasking in the
context of a wave equation solver (ExaHyPE) using three different architectures and
runtimes. We describe several task-scheduling flaws present in currently
available runtimes, demonstrate how they impact performance and show how to
work around them. Finally, we propose extensions to the OpenMP standard.

\end{abstract}

\section{Introduction}

%
%
Modern high-performance computing (HPC) architectures exhibit unprecedented hardware parallelism~\cite{EuroHPC2020,exascaleroadmap}. The potential of which must be harnessed on
the software side.  As traditional loop-based parallelism and, in particular,
the bulk-synchronous (BSP) paradigm increasingly struggle to achieve this
alone, task-based programming promises to come to the programmers' rescue.
Task graphs~\cite{4553700} allow the programmer to oversubscribe
the system logically, i.e.~to write software with a significantly higher
concurrency than the hardware provides.
Once task graphs are translated to task-based source code, a threading runtime can efficiently map this code onto the actual
hardware, as the oversubscription provides the scheduler with the freedom to
utilise all resources. 
Despite the promise and flexibility it offers, tasking as
a low-level parallelisation paradigm, i.e.~for tiny work units~\cite{Demeshko:TaskbasedAlgorithmsAndApplications:2020},
often yields inferior performance compared to more traditional parallelisation.

%
%
Our work orbits around the second instalment of the code ExaHyPE
\cite{Software:ExaHyPE}, which can solve hyperbolic equation systems in the
first-order formulation.  We focus on its patch-based Finite Volume schemes
realising block-structured adaptive mesh refinement (AMR) \cite{Dubey:16:SAMR}.
This application requires a high degree of concurrency in order to 
smooth out the imbalances introduced by dynamic AMR, consecutive solver steps
that are drastically different in their compute characteristics, and
bandwidth limitations due to a flurry of MPI activity.
We orchestrate all that with a task formalism on top of classic domain
decomposition. Despite its complexity, the code's execution time is dominated by
bursts of similar computational steps applied to a large set of unknowns. For
example, the application of the same compute kernel to a large number of cells.
This pattern is likely archetypical for many HPC codes.

%
%
Our demonstrator code ExaHyPE uses a parallelisation strategy consisting of
MPI and OpenMP~\cite{openmp5}. We note that, with current OpenMP runtimes, our implementation that uses \emph{native} OpenMP task parallelisation yields an inferior time-to-solution compared to
plain BSP-style parallelism.  Further, the task-based formalism struggles to
compete with traditional data decomposition. Our analysis identifies two primary reasons for
this behaviour in the OpenMP runtimes and versions that we studied.

\emph{BSP task subgraphs are treated as critical paths.} If the runtime encounters an imbalance, idle times are not used to swap in
further ready tasks from the non-BSP region.  Instead, we busily wait for the
completion of the BSP subgraph.

\emph{The creation of massive numbers of ready tasks is likely to lead to the
suspension of currently active tasks.} We observe the runtime to prioritise the
execution of descendent tasks before resuming the primarily active ones ---
even if there are no dependencies.

%
%
The aforementioned shortcomings of OpenMP task runtimes lead us to
reject the hypothesis that task-based parallelism helps mitigate
load imbalances and sequential program phases introduced by classic BSP-style
parallelism. We identify causal properties and propose wrappers
around existing OpenMP calls to mitigate these flaws. The patterns studied in this work are not exclusive to ExaHyPE. The ideas presented  are therefore of broader interest for the supercomputing community working with task graphs.

%
%
The remainder of the text is organised as follows: We sketch our application in
Section \ref{section:demonstrators} with a particular emphasis on two
distinct task graphs produced by two different solver implementations. After
introducing the test platform (Section \ref{section:system}), these task
patterns are analysed. We highlight shortcomings encountered with current
runtimes, and outline tweaks to the OpenMP port of our application. This
Section \ref{section:behaviour} is the main part of our contribution. A brief
summary and an outlook (Section \ref{section:conclusion}) close the discussion.

\section{Case studies}
\label{section:demonstrators}

We use the patch-based Finite Volume (FV) solver on adaptive Cartesian meshes
\cite{Dubey:16:SAMR} that comes with the second generation of the ExaHyPE engine \cite{Reinarz:2020:ExaHyPE}.
The mesh consists of squares (dimension $d=2$) or cubes ($d=3$). Each \emph{cell}
hosts a patch of $N^d$  $d$-dimensional volumes. Each \emph{volume} carries a
piece-wise constant representation of the solution, and an additional
halo-layer of $d$-dimensional volumes surrounds each patch.

Once per time-step, our FV solvers evolve all patches in time by computing the
underlying Riemann problem with a Rusanov scheme~\cite{leveque_2002} supplemented by volumetric
source terms (right-hand side).  The Rusanov scheme requires the halo layer mentioned
above to evolve the cells at the patch boundary. After the temporal evolution, the FV solvers \emph{reduce} the maximal eigenvalue of the solution over all cells.  This eigenvalue determines the largest time-step that satisfies the CFL condition~\cite{leveque_2002}.
Before the next time-step, each patch writes $4N$ ($d=2$) or $6N^2$ ($d=3$) boundary cells into a face buffer. Each cell has its own halo. This enables us to separate the halo updates into a ``project onto faces'' epilogue
of the patch solve, and a ``write halo'' preamble to a patch update, ultimately allowing the patches to be \emph{independently} advanced in time. 

%
%


\paragraph{Solver 1: Plain BSP-style adaptive time-stepping}
Our baseline code splits the computational domain into non-overlapping segments
along a space-filling Peano curve (SFC)~\cite{DBLP:journals/corr/Weinzierl15}.
Several adjacent segments along the SFC are deployed to each MPI rank.  Per
time-step, each rank maps its local SFC segments onto an OpenMP
\texttt{taskloop}~\cite{openmp5}.  The programming paradigm is classic SPMD on
the MPI side, followed by a BSP-style traversal of the local subdomains per
rank. Equivalent code using a \texttt{parallel for} construct with dynamic scheduling yields the same performance.

\begin{figure}[!htb]
    \begin{subfigure}{.5\textwidth}
      \centering
      \includegraphics[height=.3\textheight]{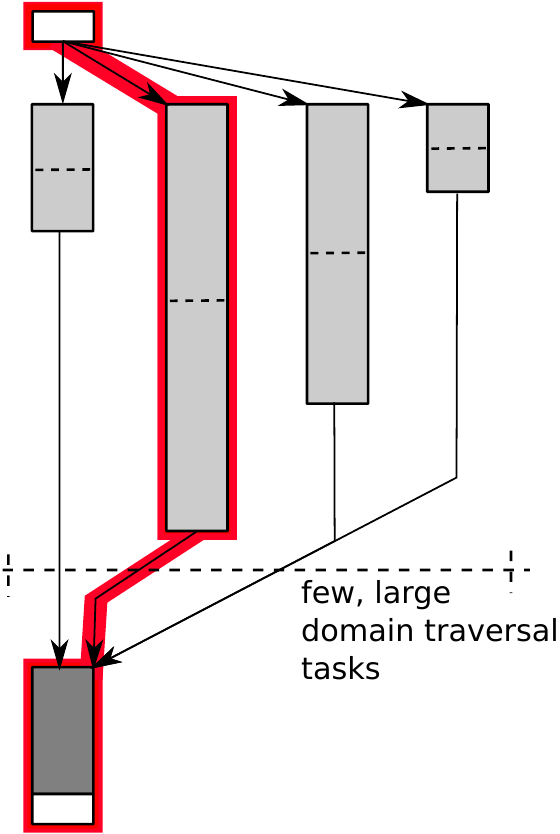}\\
      \caption{BSP-type solver}
      \label{fig:sub-firsta}
    \end{subfigure}
    \begin{subfigure}{.5\textwidth}
      \centering
      \includegraphics[height=.3\textheight]{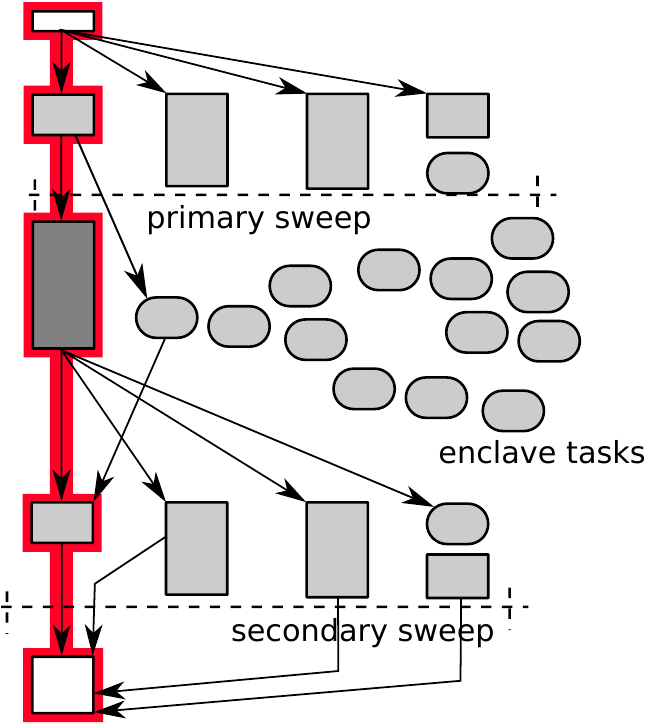}\\
      \caption{Enclave solver}
    \label{fig:sub-seconda}
    \end{subfigure}
    \caption{Anatomy of a time-step. \subref{fig:sub-firsta}): Plain domain decomposition yields one BSP-type graph per MPI rank with
  a warm up phase per time-step (white) where we determine the time-step size and global variables.
  Each local SFC subpartition is processed by one thread (light grey).
  A quasi-serial bit of code handles the MPI boundary exchange (dark grey).
  \subref{fig:sub-seconda}): In our enclave solver, the rank-local domain traversal is split into
  two BSP-type traversals.
  The first bulk produces small enclave tasks (rounded corner), while skipping local computations within the subdomain.
  It is faster than the plain counterpart and immediately triggers the boundary data exchange.
  Once completed, it spawns another bulk which weaves in the enclave task
  outcomes before completing the time-step.
  The producer-consumer dependencies are shown for one enclave task only. The critical path is highlighted in red.
  \label{figure:task-graph:enclave}
 }
\end{figure}

Every time the task encounters a cell, it writes the halo, updates the cell,
determines the next permissible time-step size, and stores the new halo data in
the faces.  The computation of the permissible time-step size is a per-cell
operation and is therefore fused with the actual cell update, meaning that one
large BSP section per time-step is sufficient.

The underlying load balancing problem is a chains-on-chains problem
\cite{Pinar:2004:ChainsOnChains}.
We operate with an SFC subpartition count that is close or equal to the number of available cores.
Due to the AMR, boundary handling and administrative overhead, the individual
subpartitions are not perfectly balanced per thread.  We obtain a classic
BSP-style task graph (\figref{figure:task-graph:enclave}), where the bulk
is not perfectly balanced. 

\paragraph{Solver 2: Enclave tasking}
%
%
%
In contrast to the plain domain decomposition scheme, this solver traverses the
mesh twice per time-step, and it classifies cells as either skeleton (those
adjacent to partition boundaries or AMR resolution transitions) or enclave
cells (all others)~\cite{Charrier:2020:EnclaveTasking}.  As soon as the primary
traversal encounters a skeleton cell, it updates it and thereby determines the
new local solution, the permissible time-step size and the new value as
required by adjacent cells in the next time-step.  This data has to be
interpolated or restricted, sent via MPI, or copied over locally to another
logical subpartition.

If the thread-local traversal hits an enclave cell, it maps this local cell
update onto a task.  At the end of this primary grid sweep, we exchange the
partition boundary data.  The secondary grid traversal waits for its task to
terminate, weaves the task outcome into the solution representation, and
reduces the permissible time-step size per rank. A final, brief serial phase
launches the global time-step size reduction and finalises all MPI data
exchanges.  

\begin{algorithm}[htb]
 \caption{
  Schematic layout of the time-stepping in our enclave tasking.
  \label{algorithm:enclaves}  
 }
 \begin{algorithmic}[1]
  \Function{timeStep}{$dt$}
   \State \texttt{\#pragma omp taskloop nogroup}  
   \For{rank-local partition}
    \Comment{Primary traversal (large task)}
    \For{local cell}
     \If{cell is skeleton}
      \State update cell
     \Else
      \State \texttt{\#pragma omp task}
       \Comment{Spawn enclave task}
      \State update cell
     \EndIf
    \EndFor
   \EndFor
   \State \texttt{\#pragma omp taskwait}
    \Comment{Wait only for traversal tasks}  
   \State Realise domain boundary exchange
   \State \texttt{\#pragma omp taskloop nogroup}  
   \For{rank-local partition}
    \Comment{Secondary traversal (large task)}
    \For{local cell}
     \If{cell is enclave}
      \State busy-wait for enclave task outcome
      \Comment{With \texttt{taskyield}}
     \EndIf
    \EndFor
   \EndFor
   \State \texttt{\#pragma omp taskwait}
    \Comment{Implicitly wait for all tasks}  
  \EndFunction
 \end{algorithmic}
\end{algorithm}

Our enclave tasking (\figref{figure:task-graph:enclave}) is realised as a
sequence of two \texttt{taskloop} constructs per time-step (Alg.~\ref{algorithm:enclaves}).
In contrast to the plain implementation, the first \texttt{taskloop} acts as a producer of
tasks. It does not synchronise with the spawned enclave tasks,
as the \texttt{nogroup} first eliminates all implicit barriers 
and the \texttt{taskwait} then waits for the direct children of
the master thread only. Despite the elimination of this barrier we continue to refer to this as task group.
The local domain decomposition remains invariant and
generates only a few relatively large tasks. The stark contrast to the plain
variant is that we create a plethora of tiny enclave tasks per primary sweep.

We use a simple hashmap for the bookkeeping of our task outcomes: Enclave tasks
are assigned a unique number at the point of their creation. Upon completion,
they reduce their permissible time-step size and enter the new time-step data
of their associated patch in the hash map.

The busy waits in the secondary traversals of Algorithm~\ref{algorithm:enclaves} incorporate simulation outcomes into the computational
mesh. This is required to allow updating the patch halo and synchronising the
patches with their neighbours.  In our baseline OpenMP implementation, this is
realised by busy waiting: The code polls the hash map repeatedly.  As long as the hash map does not yet contain the required task
outcome, the polling code releases the semaphore and issues a
\texttt{taskyield} before it polls again.  This constitutes a na\"ive
implementation of the consumer in a producer-consumer pattern.

\section{Test environment}
\label{section:system}


\begin{table}
 \caption{
   Test systems.
   \label{tab:machines}
  }
 \centering
 {\footnotesize 
 \begin{tabular}{l|c|c|c}
   \toprule
   Test system & Hamilton & HPE Hawk & Cosma \\
   \midrule
   CPU & 
     Intel Xeon E5-2650V4 & 
     AMD EPYC 7742 & 
     Intel Xeon Gold 5218 \\
   Name & 
     Broadwell & Rome & Cascade Lake \\
   Cores & 
     $2 \times 14 $ & 
     $2 \times 64 $ & 
     $2 \times 16$ \\
   NUMA domains &
     2 &
     $2 \times 4$ &
     2 \\
   Baseline freq. & 
     2.4 GHz &
     2.25 GHz &
     2.3 GHz \\
   L2/L3 & 
     256 kB/30 MB &
     512 kB/16 MB &
     1 MB/22 MB \\
   Compiler & 
     icpc (ICC) 19.1.3.304 & 
     g++ (GCC) 10.2.0 & 
     icpx (ICX) 2021.1 Beta \\
   \bottomrule
 \end{tabular}
 }
\end{table}

We work with three different test systems (\tabref{tab:machines}), each
using a different compiler: A GNU compiler, the Intel compiler, and the new
LLVM-based Intel compiler. This allows use to make qualitative statements. \emph{Quantitative}
comparisons are beyond the scope of this work.
All tasking code is based on \omp.
\texttt{OMP\_PROC\_BIND} is set to \texttt{close}.

We simulate compressible Euler equations in a unit-square domain with periodic
boundaries, hosting 59,049 patches. An initial high-density peak in the domain
serves as a causal agent for spreading waves.  We use FV with a patch size of
$63\times 63$ per cell such that the face count (Riemann problems) per patch
along each coordinate axis equals a power of two. The computational simplicity
of Euler equations implies that the code is never compute-bound. 
Due to this characteristic,
scheduling flaws become apparent immediately.

We limit our experiments to a single node but explicitly keep all
management code for inter-node data exchange
enabled, i.e.~after each time-step, we run the routines that
orchestrate multi-node and heterogeneous runs. Doing so eliminates interference with MPI.
We further disable dynamic load balancing (cf.~\cite{10.1007/978-3-030-58144-2_9}) and instead rely on two static ways
to split the domain along the Peano SFCs: For our first, \emph{well-balanced}
mode, we ensure that each core obtains one SFC-partition of the domain and that
the partition sizes (cell counts) do not differ by more than 10\%. When using a single core we end up with 58,564 enclave cells. Splitting the same domain over 24 cores yields a smaller \emph{total} of 53583 enclave cells as we have more boundaries and therefore more skeletons.  In our second,
\emph{ill-balanced} mode, we assign around half of the partition (26,244 cells)
to the first core and then continue to iteratively cut the partition size in
half for the remaining cores. Doing so yields a highly ill-balanced data
decomposition with up to 20 partitions where the smallest partition consists of
only one patch.

\section{Benchmarking and task runtime modifications}
\label{section:behaviour}

\begin{figure}[!h]
    \begin{subfigure}{.5\textwidth}
  \centering
  \includegraphics[width=0.99\textwidth]{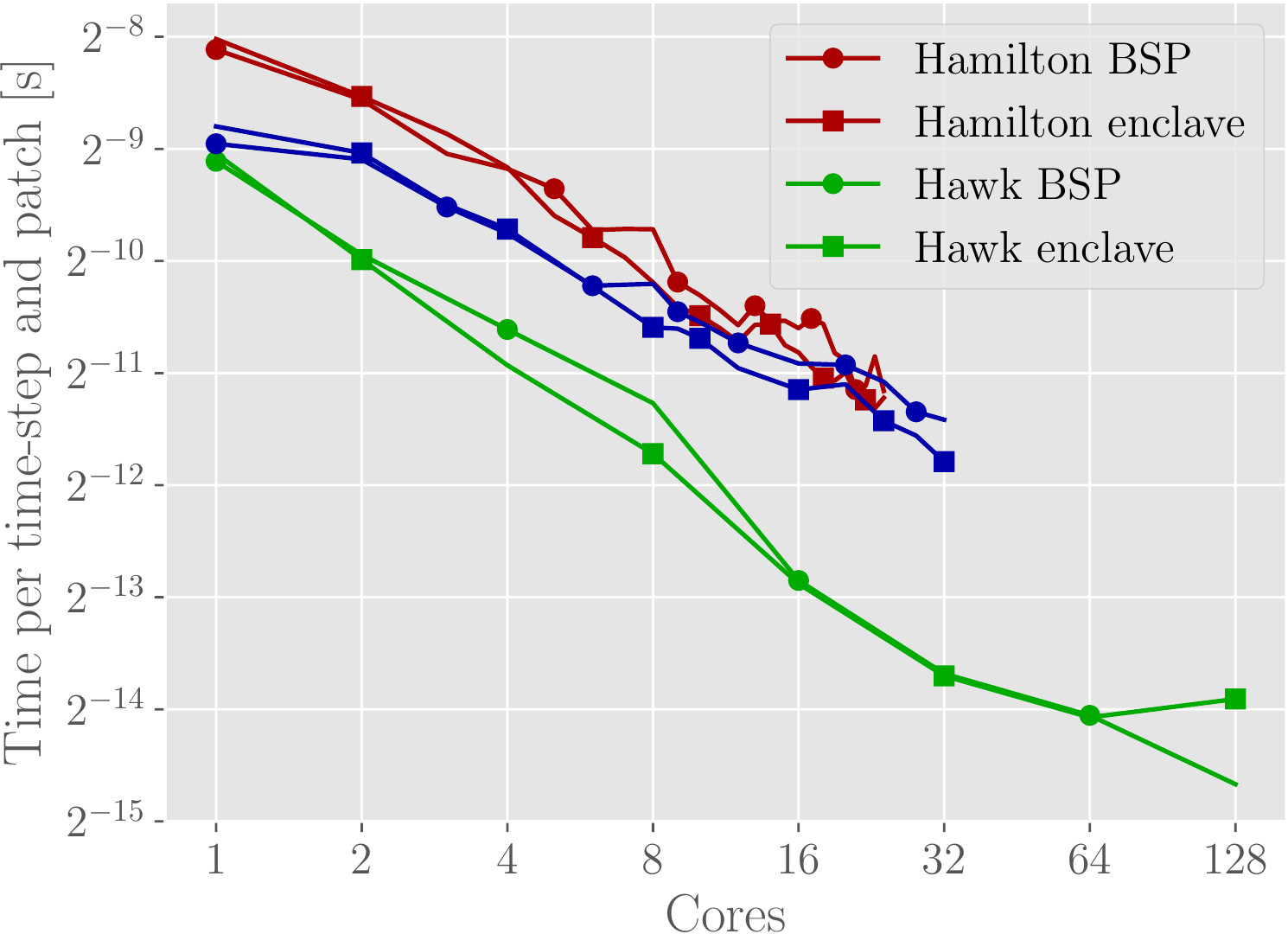}
  \caption{Well-balanced setup}
  \label{fig:scale-base-well}
\end{subfigure}
\begin{subfigure}{.5\textwidth}
  \centering
  \includegraphics[width=0.99\textwidth]{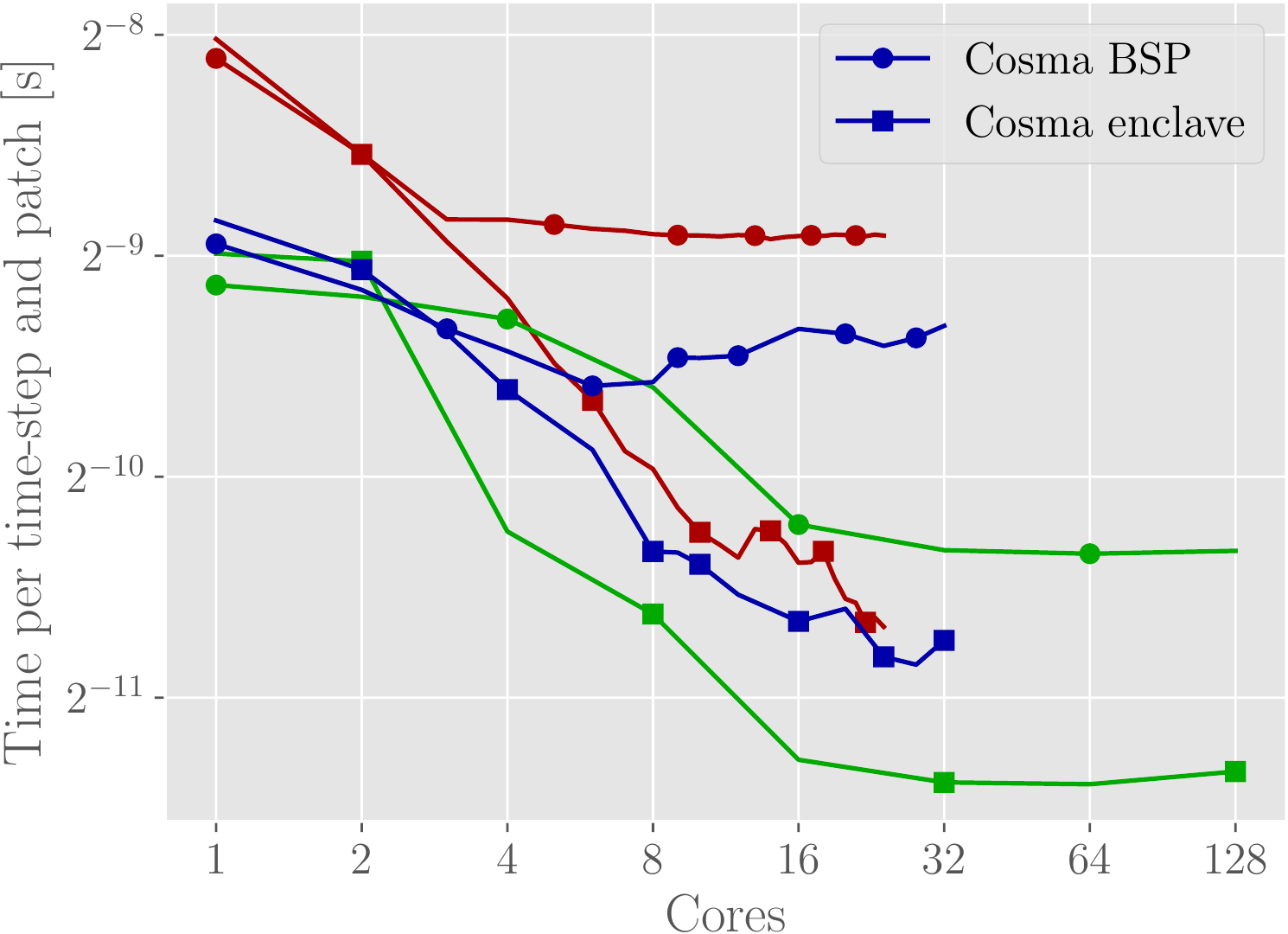}
  \caption{Ill-balanced setup}
  \label{fig:scale-base-ill}
\end{subfigure}
 \caption{
     Baseline scaling, measuring the time per time-step and patch. 
  Both the BSP and the enclave implementation use native OpenMP tasks.
  \label{figure:results:baseline}
 }
\end{figure}

%
%
We first compare the performance of the BSP-style solver and the enclave solver \emph{without}
modifications of the \omp runtime (referred to as ``native'' in the following).
The baseline BSP-type code scales robustly only for the well-balanced domain
decomposition (\figref{fig:scale-base-well}). In the ill-balanced
setup, BSP parallelisation (circles in \figref{fig:scale-base-ill}) maximally achieves a 2x-speedup (w.r.t. single-core),
regardless of the number of cores and type of machine studied.  The enclave
algorithm arguably performs better than pure BSP.
All measurements report time per time-step and patch.

%
%
The strength of enclave tasking is its ability to migrate computational work to
underutilised cores. This migration pays off for an ill-balanced setup, but its
impact in a well-balanced setup is small, as this (experimental) choice does not
benefit from the flexibility of OpenMP tasking. Overall, the strong scaling
regime exhibits limited efficiency which is, however, beneficial for our purposes as it makes
all scheduling flaws immediately apparent.

%
%

\subsection{Direct translation of enclave tasking to OpenMP (native)}

\paragraph{Busy polling}
In our baseline code, we map enclave tasks directly onto OpenMP tasks and
realise the busy-wait in Algorithm~\ref{algorithm:enclaves} via polling: We
check whether the task outcome is available and otherwise invoke
\texttt{taskyield}.  This implementation notoriously causes OpenMP runtimes to
starve once the number of domain partitions exceeds the number of OpenMP
threads: Those consumer threads of the \texttt{taskgroup} which have not yet
hit the implicit barrier take turns waking up each other instead of an enclave
task, thereby starving the latter.

\begin{observation}
    \texttt{taskyield} tends to switch between tasks within the same group. This pattern starves ready tasks outside the task loop. The OpenMP implementations tested are not "fair".
\end{observation}

\noindent
To be formally correct, our implementation should introduce dependencies between the
tasks instead of polling. We refrain from doing so as the additional bookkeeping
would require complex rewrites over multiple classes distributed among multiple components.
Some parallel algorithms, like our enclave tasking, are ``starvation-free'' yet require ``fair'' task scheduling to progress. 
If \texttt{taskyield} does not yield to other taskgroups, progress for these algorithms can be minimal or the code can starve.
In contrast, non-fair scheduling switching between few ready tasks is advantageous in many situations as it avoids cache capacity misses.

\begin{feature}
    It is desirable to annotate \texttt{taskyields} with scheduling hints to allow it to also process tasks that are \emph{not direct} descendants.
\end{feature}

\noindent
Since we lack a cross-taskgroup yield, oversubscribing the machine threads with
large traversal tasks is not an option. The flexibility that geometric load
balancing offers is therefore limited~\cite{10.1007/978-3-030-58144-2_9}.  More flexible scheduling
(\texttt{KMP\_TASK\_STEALING\_CONSTRAINT}) or untied task progression do not
help in this situation: They facillitate optimisation \emph{within} a given scheduling strategy (cf.~\cite{10.1007/978-3-319-98521-3_1} for a discussion on \texttt{taskyield} behaviour) but do not allow to \emph{change} the strategy itself.


\begin{figure}
    \begin{subfigure}{.5\textwidth}
  \centering
  \includegraphics[width=0.95\textwidth]{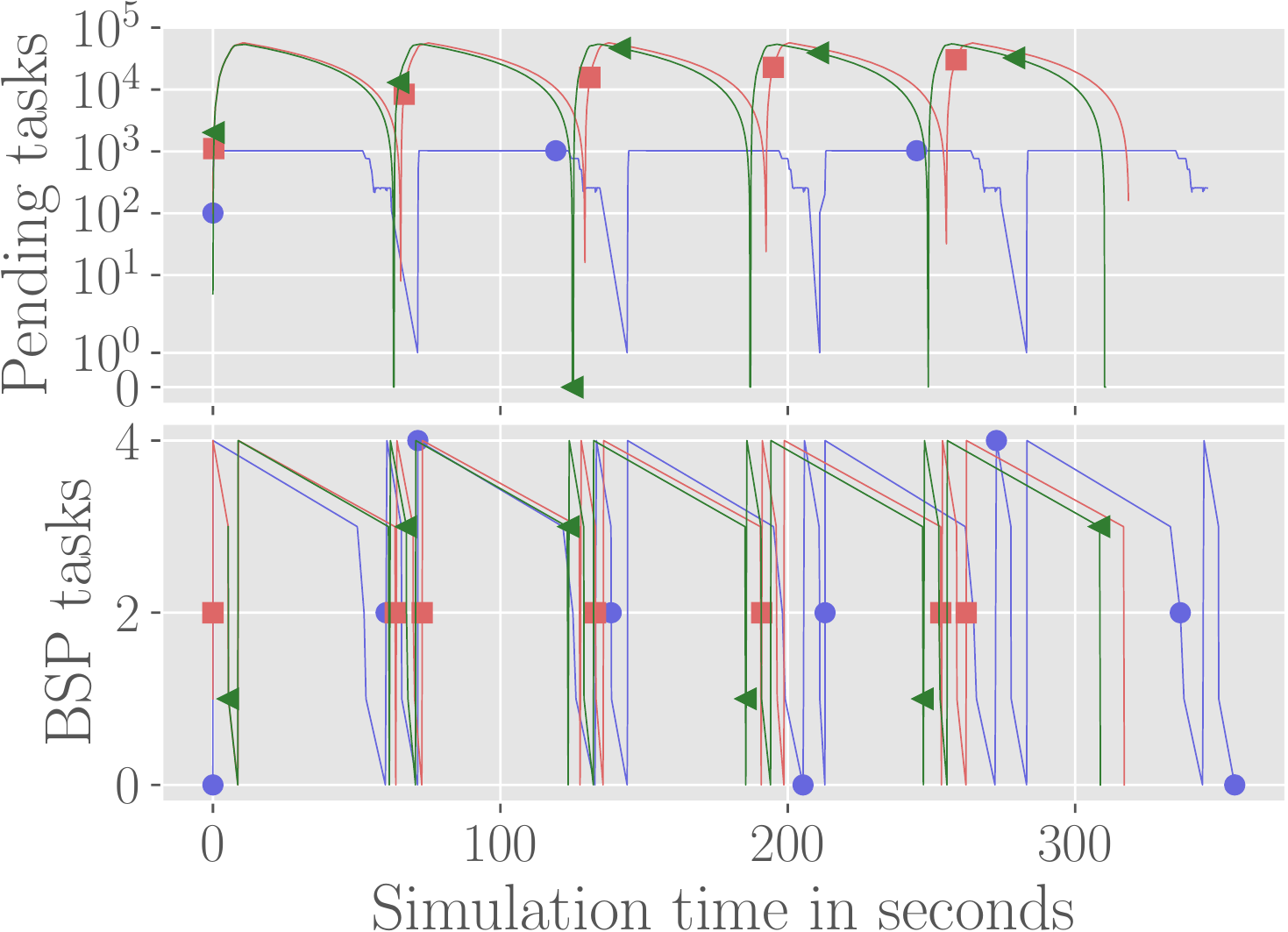}
  \\[.4em]
  \includegraphics[width=0.95\textwidth]{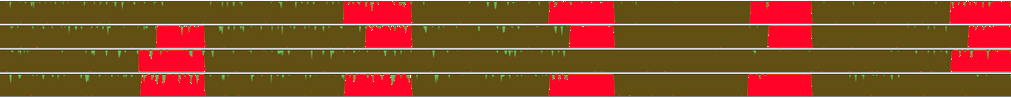}
  \caption{Well-balanced setup}
  \label{fig:sub-first}
\end{subfigure}
\begin{subfigure}{.5\textwidth}
  \centering
  \includegraphics[width=0.95\textwidth]{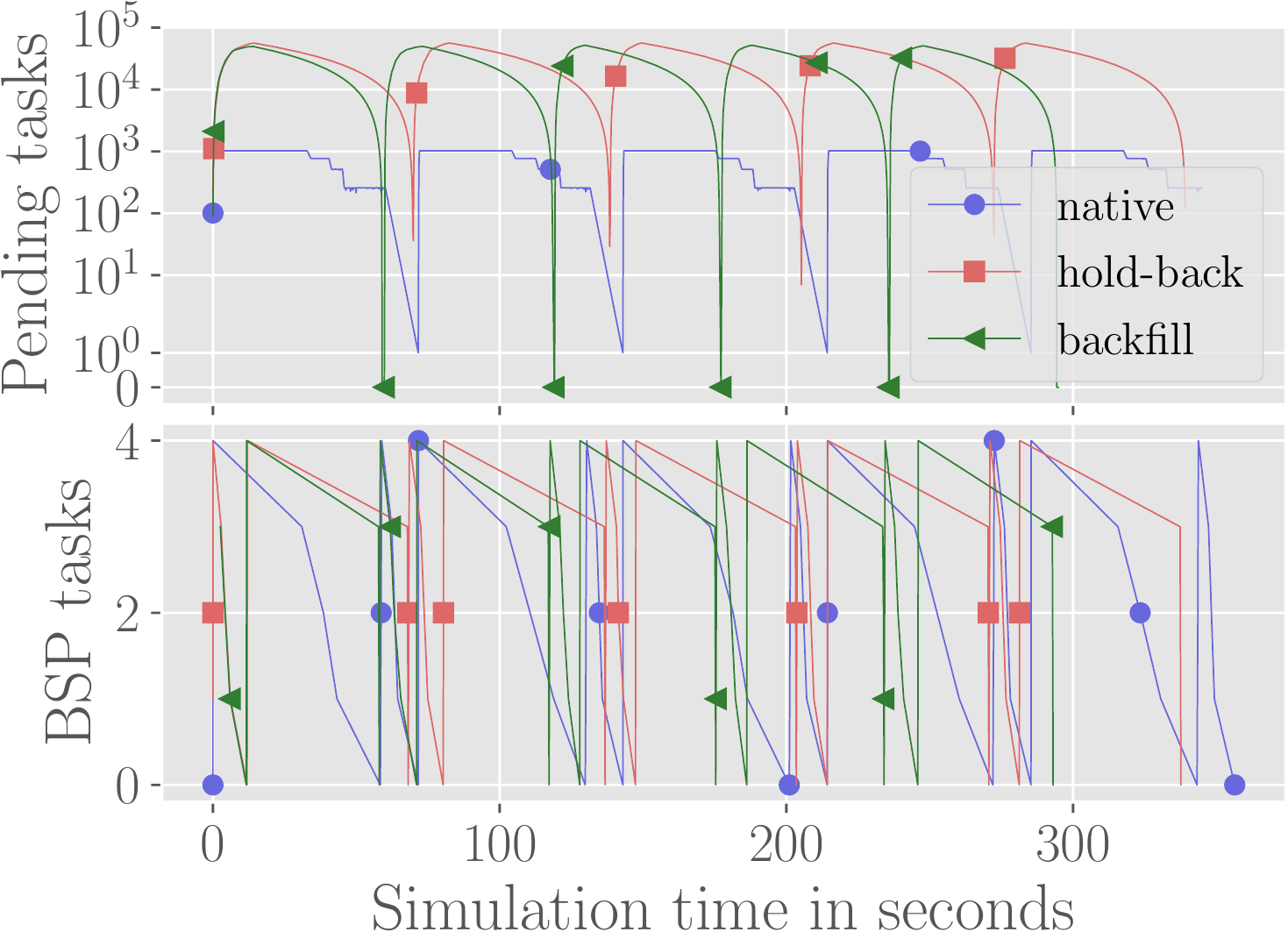}
  \\[.4em]
  \includegraphics[width=0.95\textwidth]{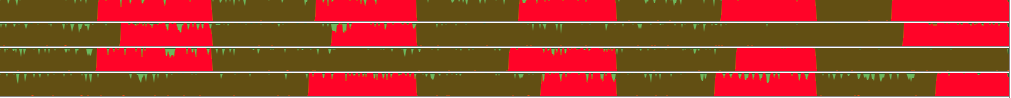}
  \caption{Ill-balanced setup}
  \label{fig:sub-second}
\end{subfigure}
 \caption{
     Analysis of five time-steps.
  Top:
  Number of pending enclave tasks for three different tasking realisations. The native OpenMP implementation prevents spawning more than 1,000 enclave tasks.
  Middle:
  Number of active BSP (producer/consumer) tasks. The two traversals per time-step are clearly visible. The native OpenMP implementation leads to the first traversal taking much longer than the second. For the other implementations, the behaviour is the opposite.
  Bottom:
  Core activities of the native implementation (brown: CPU time, red: spinning).
  The data show five time-steps of two runs on Hamilton utilising four cores.
  \label{figure:plain-omp:pending-task-statistics}
 }
\end{figure}


\paragraph{Task execution pattern}

For the enclave tasking to have its desired effect, it is imperative to spawn a
large number of tasks during the first traversal to have them pending for the
\emph{second} traversal. Our measurements (\figref{figure:plain-omp:pending-task-statistics}) clearly show that
the native OpenMP implementation caps the number of tasks at about 1,000. The
runtime processes pending tasks immediately. In doing so, the completion of the
majority of the tasks coincides with the \texttt{taskwait} clause at the end of
the primary traversal (Alg.~\ref{algorithm:enclaves}, line 13). All our OpenMP
implementations use this scheduling point to process a significant amount of enclave tasks (ready by
definition).  Once the runtime progresses beyond the synchronisation point, in
the native implementation, only a few enclave tasks (if any) remain. This
implies that any idle time in the subsequent BSP-type traversal cannot be
backfilled with further ready tasks. The resulting wait at the end of the
secondary sweep causes the spin times in the traces measured.  This behaviour
leads to the first traversal lasting significantly longer than the second one,
exactly the opposite of what enclave tasking requires.  It should be noted that
an ill-balanced domain partitioning amplifies the observed patterns.  The
following paragraphs describe how we circumvent these effects.

\subsection{Manual task postponing (hold-back)}

%
%
Thresholds that influence the task processing behaviour degrade the
predictability of the runtime performance, and they destroy our code's
efficiency: We create enclave tasks to compensate for imbalances once the
primary taskloop has terminated.  If OpenMP decides to suspend the task
producer and instead launches tasks immediately, or OpenMP does not continue
with the main control flow at the synchronisation point, the desired pay-off
vanishes.

\begin{observation}
OpenMP runtimes switch to immediate task processing if the number of ready
tasks exceeds a threshold.  This behaviour introduces the risk that these
tasks are not available to mend subsequent imbalances.  It also denies the
programmer the opportunity to migrate expensive but not immediately
required work to subtasks.
\end{observation}

\begin{observation}
 OpenMP's \texttt{taskwait} allows the runtime to switch to processing any ready tasks---not only the child tasks of the current task region---rather than continuing with the main control flow. That suspends the task producer thread.
\end{observation}

\begin{figure}[!h]
    \begin{subfigure}{.5\textwidth}
  \centering
  \includegraphics[width=0.99\textwidth]{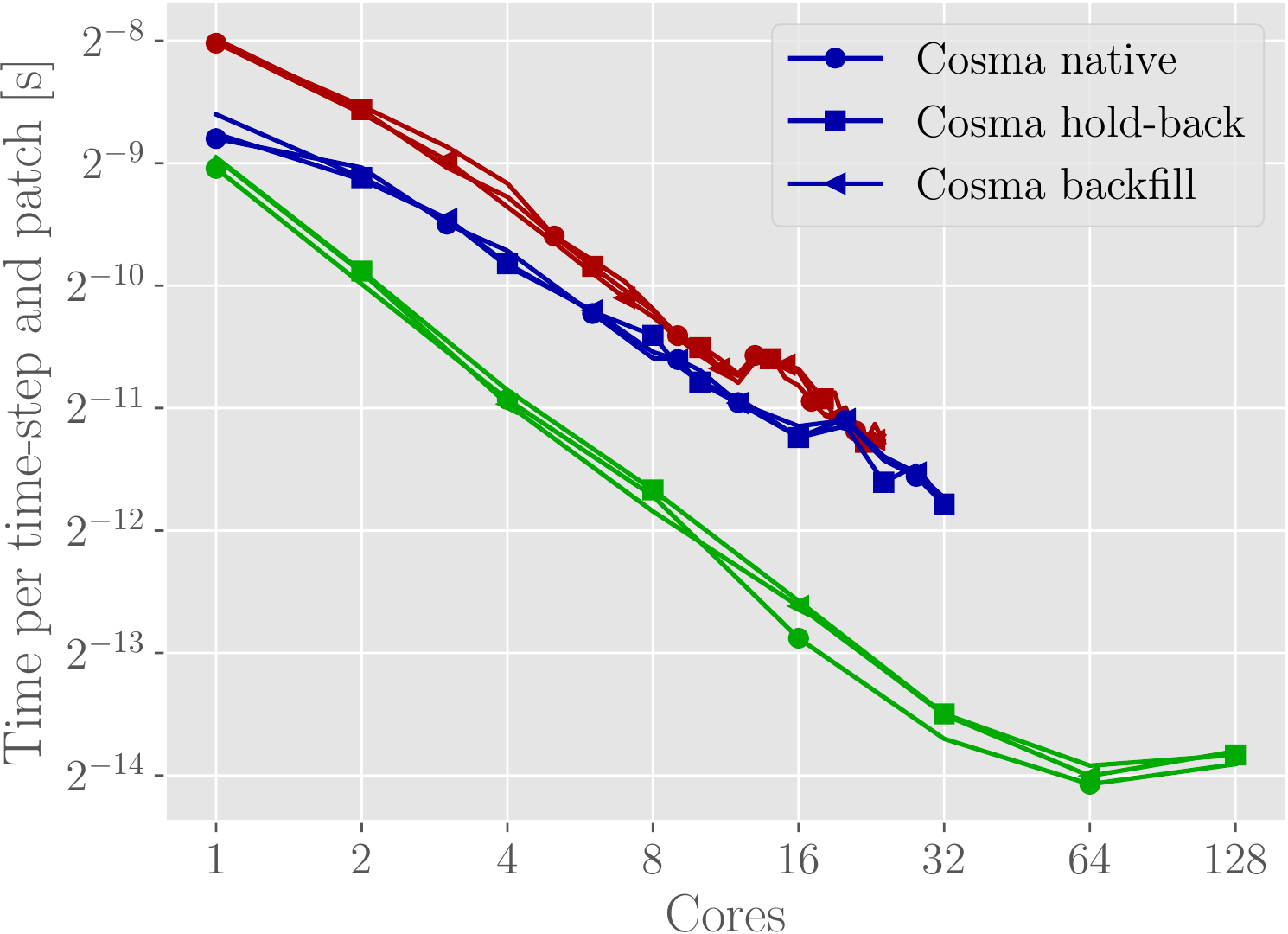}
  \caption{Well-balanced setup}
  \label{fig:scale-well}
\end{subfigure}
\begin{subfigure}{.5\textwidth}
  \centering
  \includegraphics[width=0.99\textwidth]{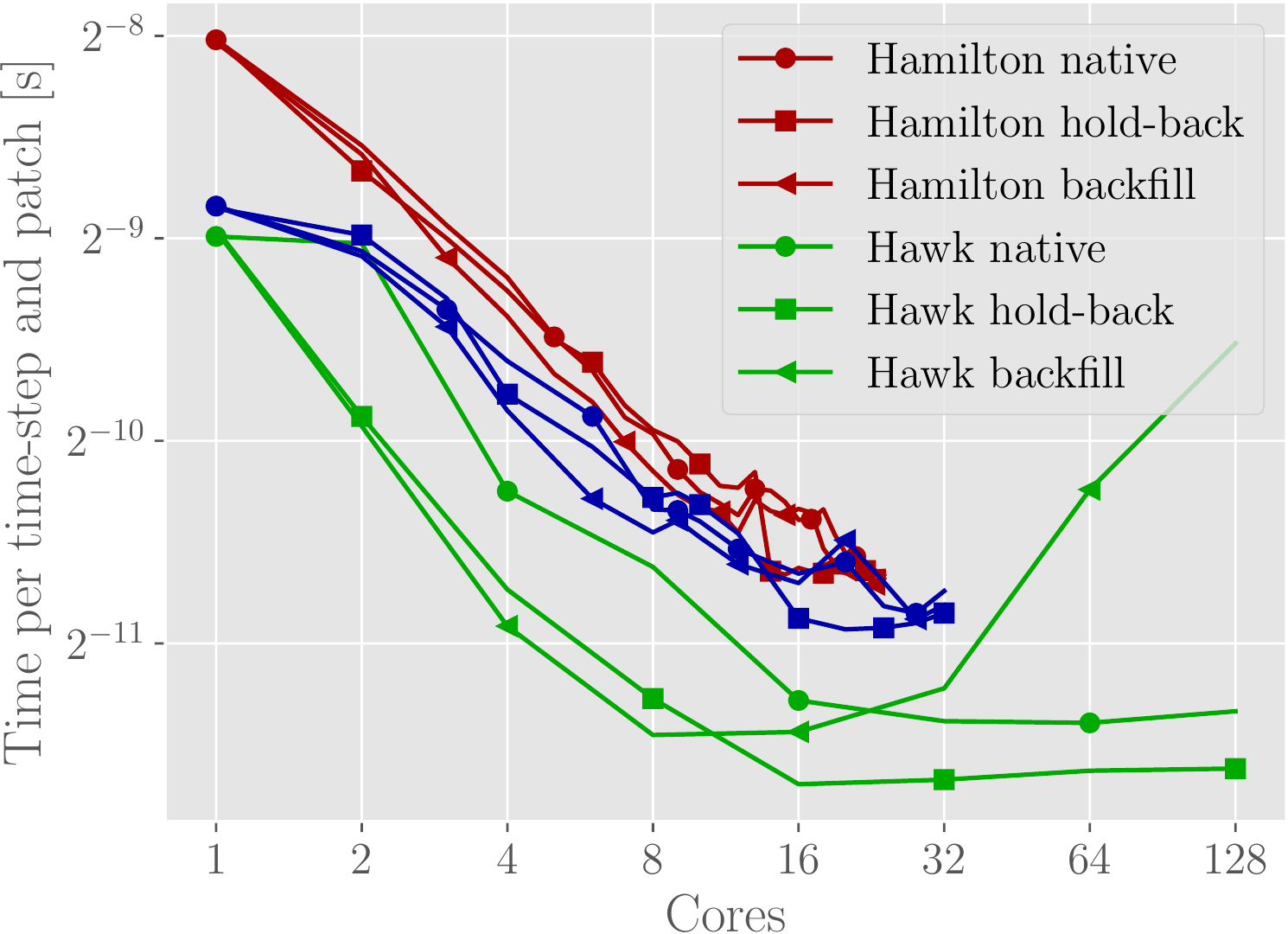}
  \caption{Ill-balanced setup}
  \label{fig:scale-ill}
\end{subfigure}
 \caption{
  Three different enclave tasking strategies per machine: Native mapping onto
  OpenMP tasks, manually hold-back of spawned enclave tasks, and a manual
  backfilling of idling cores. We measure the time per time-step and patch. 
  \label{figure:results:runtimes}
 }
\end{figure}

%
%

\noindent
Our code eliminates the immediate task processing by
adding a manual queue: 
Instead of spawning OpenMP tasks, the BSP-type task
regions queue their tasks in this helper container.
They \emph{hold back} the tasks manually.
The busy polling checks whether a task outcome is available, otherwise
processes one task from our manual queue, and then checks again.
It implicitly realises a lazy task evaluation.

%
%
Our measurements show that such an additional, thin, user-defined tasking layer on top of OpenMP ensures
that all enclave tasks remain pending while we continue to spawn tasks (\figref{figure:plain-omp:pending-task-statistics}).
In contrast to the native implementation, we now see (\figref{figure:plain-omp:pending-task-statistics}, middle panel plots) that the primary traversal
is very short compared to its secondary counterpart.  The reason is that the
manual queue \emph{collects} all tasks instead of processing them.  Consequently, the
secondary traversal dominates the runtime thanks to the lazy evaluation.  The
secondary sweep's polling does not discriminate which tasks are spawned by
which BSP-type task. It simply grabs the tasks one by one and automatically
balances out the secondary BSP tasks.  This advantageous behaviour is much more
pronounced in the ill-balanced setup.

\begin{feature}
It is desirable to manually control OpenMP's ready task thresholds or
inform the runtime that many tasks will be spawned, and although they are
ready, not to process them right away.
\end{feature}

\begin{feature}
It is desirable to manually annotate OpenMP's scheduling points that the
(serial) control flow in the code is part of the critical path. 
\end{feature}

\noindent
It is reasonable to introduce a threshold for ready tasks to avoid excessive
bookkeping overhead incurred by long task queues.  However, we have an
algorithm that suffers tremendously from immediate task processing as it relies
on bursts of ready tasks to compensate for task ill-balancing in subsequent
computational phases.  In this case, injecting domain knowledge (``do not
process immediately'') into task scheduling reduces the time-to-solution.
Analogous reasoning holds for scheduling points: It is hugely advantageous to
inform the runtime of the program's critical path along with the control flow.
By construction, this information is unknown to OpenMP, which relies on a
dynamic assembly of the task graph.

%
%
Our modifications do not have a sizeable effect on the runtime for a
well-balanced setup. They, however, do improve the time-to-solution for an
ill-balanced setup (\figref{figure:results:runtimes}).  Unfortunately, this
improvement is not robust: It holds for low core counts and once we go beyond
one socket.  Three effects compete here: Firstly, the hold-back mechanism
avoids that OpenMP hits a synchronisation point (\texttt{taskwait} after the
primary sweep) and thereby ensures progression along the critical path.
Secondly, it reduces any cache thrashing that arises from balancing out task
workloads within the primary traversal.  Both effects bring down the runtime of
the BSP-type task production.  Thirdly, the centralised task queue increases
the coordination pressure (semaphore access) between the tasks.

\subsection{Manual backfilling (backfill)}

Manual task postponing (holding back) and the native OpenMP task processing
behaviour materialise two extreme cases of task scheduling:
They either process tasks relatively early (around the synchronisation point) or
very late due to our lazy mechanism.

It is not immediately clear which approach is more beneficial. On the one hand,
making \texttt{taskwait} work through the set of ready tasks as aggressively as
possible reduces the bookkeeping overhead, and for many codes, we may assume
that any ready task will spawn further tasks.  Discovering this early reveals
more fragments of the final task graph. This approach yields high throughput.
If, on the other hand, we make \texttt{taskwait} busy poll its siblings instead
of processing further tasks, we prevent situations where the BSP-type subgraph
terminates, but the runtime does not immediately continue with the source code
following the BSP-type section. This approach eliminates latency along the
BSP-type subgraph.

\begin{algorithm}[htb]
 \caption{
  Manual backfilling of a BSP-type task section. 
  \label{algorithm:backfill}  
 }
 \begin{algorithmic}[1]
  \State $busyThreads \gets max(\#threads,\#bsp\ tasks)$
    \Comment Ensure all threads are used
    \For{$i=0 .. busyThreads-1$} \Comment A \texttt{parallel for} would be equivalent
    \State \texttt{\#pragma omp task shared(busyThreads)}
    \State \texttt{\{}
     \If{$i<\#tasks$}
      \State \Call{run}{$task[i]$}
     \EndIf
     \State \texttt{\#pragma omp atomic}
     \State $busyThreads \gets busyThreads-1$
     \While{$busyThreads > 0 \wedge busyThreads<\#threads$}
       \Comment Second clause 
       \State \Call{processPendingTasks}{}
       \Comment avoids deadlocks 
     \EndWhile
    \State \texttt{\}}
  \EndFor
  \State \texttt{\#pragma omp taskwait}
 \end{algorithmic}
\end{algorithm}

\noindent
Our code requires a compromise between high throughput and low latency, as any
delay along the control flow with
the BSP section will introduce imbalances and delays later down the line.  We,
therefore, augment the postponed scheduling with a manual task backfilling
(Alg.~\ref{algorithm:backfill}): Enclave tasks created within the task
group (BSP-style) are enqueued using a container as before.  They are not
handed over to OpenMP.
Once a BSP-type task  terminates, it decrements a global counter of active BSP tasks
($busyThreads$).  If there are fewer active BSP-type tasks than logical
threads, and not all BSP-type tasks have terminated yet, we grab tasks from the
local task queue and process them immediately.  This is the actual backfilling, which
is in essence a conservative form of work-stealing \cite{doi:10.1177/1094342011434065}.

The backfilling ensures that our latency-sensitive BSP-subgraph realisation does not let
threads idle. Therefore, the backfilling robustly outperforms a native OpenMP
task implementation, as long as we utilise only one socket
(\figref{figure:results:runtimes} --- one socket means 14 cores for Hamilton, 16 cores for Cosma, 64 cores for Hawk).  If we have a well-balanced setup, the
backfilling does not kick in.  We, however, benefit from the payoffs of the
hold-back strategy, which automatically balances partitions with different
numbers of enclave tasks.  These tasks differ, even if the partitions all have
similar cell count.  If we have an ill-balanced setup, backfilling outperforms
a native OpenMP version, as we benefit from the hold-back mechanisms but do not
let cores idle. The benefits disappear as soon as we use both sockets or the
problem gets too small. The program suffers from cache thrashing and
synchronisation overhead, and is therefore outperformed by the hold-back
strategy.
\begin{feature}
It would be beneficial if a \texttt{taskwait} or implicit BSP
synchronisation could be annotated whether throughput or latency (immediate
continuation) take priority.
\end{feature}
\noindent
Our backfilling wraps around OpenMP's BSP constructs
(\texttt{taskloop}), and makes it latency-aware: The implementation works well if the
BSP-graph section is aligned with the task graph's critical path, and thus
latency-critical.  However, it does not prioritise latency above all else.
Instead, it tries to process enclave tasks ---
but only if other tasks are still busy with the BSP section. It is thus
\emph{weakly} latency-aware.

\section{Evaluation and conclusion}
\label{section:conclusion}

%
%
Our studies start from the observation that a plain taskification of source code with OpenMP
does not necessarily reduce the time-to-solution for sophisticated codes.  
If a code is intrinsically BSP-style, we should map it onto BSP constructs.
If we add tasking on top of these BSP regions, we quickly suffer from poor performance.
The
reason for this is not solely rooted in tasks of low arithmetic
intensity, but also stems from the fact that OpenMP runtime characteristics
impede performance as soon as we go beyond a pure tree-based task-graph layout.

%
%
We propose extensions of tasking runtimes and their API.
They can be summarised as a proposal to allow
the programmer to inform OpenMP about the criticality and characteristics of
tasks (implying statements on the arithmetic intensity and task type
homogeneity) as well as to facilitate balancing manually between throughput- and
latency-prioritisation.
Some of this information is available in approaches with a priori, i.e.~static task graph assembly
\cite{Haensel:2020:Eventify,10.1145/2929908.2929916} or can be mapped onto OpenMP's task priorities, though the latter is not fully implemented in the OpenMP runtimes that we used.
Our approach goes beyond sole prioritisation and does not require a static task graph assembly.
Instead, we wrap task APIs to include more domain knowledge about the long-term knock-on effects of scheduling decisions.
We are confident that this idea is of value for many
codes that exhibit more of a consumer-producer tasking pattern. It is worthwhile to discuss how to make these concepts available within OpenMP.

%
%
It is safe to
assume the performance gain from our techniques will be significantly higher if
the queues and scheduling are directly integrated into the runtime.
Different to state-of-the-art queue implementations, our queue is not distributed and thus suffers from congestion if many threads check it simultaneously. 
Different to high-level frameworks like
Kokkos~\cite{CarterEdwards20143202} and RAJA~\cite{rajapaper,rajasoftware}
it also lacks any affinity knowledge \cite{Terboven:2016:TaskAffinity} and thus stresses the caches.

\section*{Acknowledgements}

Holger's and Tobias' work is sponsored by EPSRC under the ExCALIBUR Phase I
call through the 
grants EP/V00154X/1 (ExaClaw) and EP/V001523/1 (Massively Parallel
Particle Hydrodynamics for Engineering and Astrophysics).
Both appreciate the support from ExCALIBUR's cross-cutting tasking theme (grant
ESA 10 CDEL).
The Exascale Computing ALgorithms \& Infrastructures Benefiting UK Research
(ExCALIBUR) programme is supported by the UKRI Strategic Priorities Fund. 
The programme is co-delivered by the Met Office on behalf of PSREs and EPSRC on behalf of UKRI partners, NERC, MRC and STFC. 
The present software \cite{Software:ExaHyPE} is part of a major rewrite of the
original ExaHyPE code funded by the European Union’s Horizon 2020
research and innovation programme under grant agreement No 671698 (ExaHyPE).
Oleksandr's work motivating this research has received funding from the
European Union’s Horizon 2020 research and innovation programme under
the project CoE POP, grant agreement No.~824080.

Our work made use of the facilities of the Hamilton HPC Service of Durham
University, and it also
made use of the facilities 
provided by the ExCALIBUR Hardware and Enabling Software programme, funded by
BEIS via STFC grants ST/V001140/1 and ST/V002724/1, and hosted by the
DiRAC@Durham Memory Intensive facility managed by the Institute for
Computational Cosmology on behalf of the STFC DiRAC HPC Facility
(www.dirac.ac.uk). The equipment was funded by BEIS capital funding via STFC
capital grants ST/P002293/1, ST/R002371/1 and ST/S002502/1, Durham University and STFC operations grant ST/R000832/1. DiRAC is part of the UK's National e-Infrastructure.

This work was funded under the embedded CSE programme of the ARCHER2 UK National Supercomputing Service (http://www.archer2.ac.uk), grant no ARCHER2-eCSE04-2.

%
%

\bibliographystyle{splncs04}
\bibliography{references}

\begin{thebibliography}{10}
\providecommand{\url}[1]{\texttt{#1}}
\providecommand{\urlprefix}{URL }
\providecommand{\doi}[1]{https://doi.org/#1}

\bibitem{EuroHPC2020}
{EuroHPC2020: EuroHPC supercomputer systems. European Commission.}
  \url{http://eurohpc.eu/} (2021)

\bibitem{4553700}
Ayguade, E., Copty, N., Duran, A., Hoeflinger, J., Lin, Y., Massaioli, F.,
  Teruel, X., Unnikrishnan, P., Zhang, G.: The design of openmp tasks. IEEE
  Transactions on Parallel and Distributed Systems  \textbf{20}(3),  404--418
  (2009). \doi{10.1109/TPDS.2008.105}

\bibitem{rajapaper}
Beckingsale, D.A., Burmark, J., Hornung, R., Jones, H., Killian, W., Kunen,
  A.J., Pearce, O., Robinson, P., Ryujin, B.S., Scogland, T.R.W.: {RAJA:}
  portable performance for large-scale scientific applications (2021), 2019
  IEEE/ACM International Workshop on Performance, Portability and Productivity
  in HPC (P3HPC)

\bibitem{Charrier:2020:EnclaveTasking}
Charrier, D., Hazelwood, B., Weinzierl, T.: Enclave tasking for dg methods on
  dynamically adaptive meshes. SIAM Journal on Scientific Computing
  \textbf{42}(3),  C69--C96 (2020)

\bibitem{Demeshko:TaskbasedAlgorithmsAndApplications:2020}
Demeshko, I., et~al.: Tbaa20: Taskbased algorithms and applications. doe report
  la-ur-21-20928 (2021),
  \url{https://permalink.lanl.gov/object/tr?what=info:lanl-repo/lareport/LA-UR-21-20928}

\bibitem{exascaleroadmap}
Dongarra, J., Beckman, P., Moore, T., Aerts, P., Aloisio, G., Andre, J.C.,
  Barkai, D., Berthou, J.Y., Boku, T., Braunschweig, B., Cappello, F., Chapman,
  B., Chi, X., Choudhary, A., Dosanjh, S., Dunning, T., Fiore, S., Geist, A.,
  Gropp, B., Yelick, K.: The international exascale software project roadmap 1.
  IJHPCA  \textbf{25},  3--60 (02 2011). \doi{10.1177/1094342010391989}

\bibitem{Dubey:16:SAMR}
Dubey, A., Almgren, A.S., Bell, J.B., Berzins, M., Brandt, S.R., Bryan, G.,
  Colella, P., Graves, D.T., Lijewski, M., L{\"{o}}ffler, F., O'Shea, B.,
  Schnetter, E., van Straalen, B., Weide, K.: A survey of high level frameworks
  in block-structured adaptive mesh refinement packages. CoRR  \textbf{74}(12),
   3217--3227 (2016)

\bibitem{CarterEdwards20143202}
Edwards, H.C., Trott, C.R., Sunderland, D.: {Kokkos:} enabling manycore
  performance portability through polymorphic memory access patterns. Journal
  of Parallel and Distributed Computing  \textbf{74}(12),  3202 -- 3216 (2014).
  \doi{10.1016/j.jpdc.2014.07.003},
  \url{http://www.sciencedirect.com/science/article/pii/S0743731514001257},
  domain-Specific Languages and High-Level Frameworks for High-Performance
  Computing

\bibitem{Haensel:2020:Eventify}
Haensel, D., Morgenstern, L., Beckmann, A., Kabadshow, I., Dachsel, H.:
  Eventify: Event-based task parallelism for strong scaling. In: Proceedings of
  the Platform for Advanced Scientific Computing Conference (2020)

\bibitem{leveque_2002}
LeVeque, R.J.: Finite Volume Methods for Hyperbolic Problems. Cambridge Texts
  in Applied Mathematics, Cambridge University Press (2002).
  \doi{10.1017/CBO9780511791253}

\bibitem{doi:10.1177/1094342011434065}
Olivier, S.L., Porterfield, A.K., Wheeler, K.B., Spiegel, M., Prins, J.F.:
  Openmp task scheduling strategies for multicore numa systems. The
  International Journal of High Performance Computing Applications
  \textbf{26}(2),  110--124 (2012). \doi{10.1177/1094342011434065},
  \url{https://doi.org/10.1177/1094342011434065}

\bibitem{openmp5}
{OpenMP Architecture Review Board}: {OpenMP} application program interface
  version 5.0 (Nov 2018),
  \url{https://www.openmp.org/wp-content/uploads/OpenMP-API-Specification-5.0.pdf}

\bibitem{10.1007/978-3-030-58144-2_9}
Orland, F., Terboven, C.: A case study on addressing complex load imbalance in
  openmp. In: Milfeld, K., de~Supinski, B.R., Koesterke, L., Klinkenberg, J.
  (eds.) OpenMP: Portable Multi-Level Parallelism on Modern Systems. pp.
  130--145. Springer International Publishing, Cham (2020)

\bibitem{Pinar:2004:ChainsOnChains}
Pinar, A., Aykanat, C.: Fast optimal load balancing algorithms for
  1dpartitioning. Journal of Parallel and Distributed Computing
  \textbf{64}(8),  974--996 (2004)

\bibitem{rajasoftware}
RAJA: {RAJA} performance portability layer. (2021),
  \url{https://github.com/LLNL/RAJA}

\bibitem{Reinarz:2020:ExaHyPE}
Reinarz, A., et~al.: {ExaHyPE}: An engine for parallel dynamically adaptive
  simulations of wave problems. Computer Physics Communications  \textbf{254},
  107251 (2020)

\bibitem{10.1145/2929908.2929916}
Schaller, M., Gonnet, P., Chalk, A.B.G., Draper, P.W.: Swift: Using task-based
  parallelism, fully asynchronous communication, and graph partition-based
  domain decomposition for strong scaling on more than 100,000 cores. In:
  Proceedings of the Platform for Advanced Scientific Computing Conference.
  PASC '16, Association for Computing Machinery, New York, NY, USA (2016).
  \doi{10.1145/2929908.2929916}, \url{https://doi.org/10.1145/2929908.2929916}

\bibitem{10.1007/978-3-319-98521-3_1}
Schuchart, J., Tsugane, K., Gracia, J., Sato, M.: The impact of taskyield on
  the design of tasks communicating through mpi. In: de~Supinski, B.R.,
  Valero-Lara, P., Martorell, X., Mateo~Bellido, S., Labarta, J. (eds.)
  Evolving OpenMP for Evolving Architectures. pp. 3--17. Springer International
  Publishing, Cham (2018)

\bibitem{Terboven:2016:TaskAffinity}
Terboven, C., Hahnfeld, J., Teruel, X., Mateo, S., Duran, A., Klemm, M.,
  Olivier, S.L., de~Supinski, B.R.: Approaches for task affinity in openmp. In:
  Maruyama, N., de~Supinski, B.R., Wahib, M. (eds.) OpenMP: Memory, Devices,
  and Tasks. pp. 102--115. Springer International Publishing, Cham (2016)

\bibitem{Software:ExaHyPE}
Weinzierl, T., et~al.: {ExaHyPE}---an exascale hyperbolic {PDE} engine (2021),
  \url{http://www.exahype.eu}, www.exahype.eu

\bibitem{DBLP:journals/corr/Weinzierl15}
Weinzierl, T.: The peano software - parallel, automaton-based, dynamically
  adaptive grid traversals. CoRR  \textbf{abs/1506.04496} (2015),
  \url{http://arxiv.org/abs/1506.04496}

\end{thebibliography}
\appendix

\section{Experiments}

All experiments rely on the sole ExaHyPE code base (fourth generation branch \texttt{p4}) which can be obtained freely through 
\begin{verbatim}
git clone -b p4 https://gitlab.lrz.de/hpcsoftware/Peano.git
\end{verbatim}

\noindent
{\bf icpc.} For the Intel/Hamilton experiments, we configure the code base via
\begin{verbatim}
./configure CXX=icpc OPENMP_LDFLAGS="-qopenmp" CXXFLAGS="-O3 \
 -xhost -std=c++14 -qopenmp" --with-multithreading=omp \
 --enable-exahype --enable-loadbalancing-toolbox
\end{verbatim}

\noindent
{\bf gcc.} For the GCC/Hawk experiments, we configure the code base via
\begin{verbatim}
./configure CXX=g++ OPENMP_LDFLAGS="-fopenmp" CXXFLAGS=" -Ofast \
 -march=native -std=c++14 -fopenmp" --with-multithreading=omp \
 --enable-exahype --enable-loadbalancing-toolbox
\end{verbatim}

\noindent
{\bf icpx.} For the Intel/Cosma experiments, we configure the code base via
\begin{verbatim}
./configure CXX=icpx CXXFLAGS="-O3 -xhost -std=c++14 -qopenmp" \
 OPENMP_LDFLAGS="-qopenmp"  --with-multithreading=omp 
 --with-multithreading=omp \
 --enable-exahype --enable-loadbalancing-toolbox
\end{verbatim}

\noindent
A successful configure call yields makefiles such that Peano's core libraries can be compiled.
As we add ExaHyPE's enable command, all ExaHyPE core libraries are built, too.

To rebuild the executables, we use a Python script
\texttt{create-paper-executables.py}.
It corresponds to a call of the Python scripts within Peano's ExaHyPE exmaple
directories of the type

\begin{verbatim}
python3 example-scripts/finitevolumes-with-ExaHyPE2-benchmark.py \
 -cs 0.005 -et 0.0001 -t default-ats -ps 63
python3 example-scripts/finitevolumes-with-ExaHyPE2-benchmark.py \
 -cs 0.005 -et 0.0001 -t enclave-ats -ps 63
python3 example-scripts/finitevolumes-with-ExaHyPE2-benchmark.py \
 -cs 0.005 -et 0.001  -t enclave-ots -ps 63
\end{verbatim}

\noindent
but only offers the setups as discussed in the paper.
It also can replace the default load balancing scheme with an (artificially)
ill-balanced one.

The actual program runs are triggered through
\begin{verbatim}
./peano4 --threading-model x
\end{verbatim}

\noindent
where \texttt{x} is either \texttt{native} (to map all tasks directly onto OpenMP tasks), 
\linebreak
\texttt{hold-back} (queue in an auxiliary data structure and map onto OpenMP tasks after
large tasks have terminated), \texttt{backfill} (use some tasks from auxiliary queue to 
manually keep imbalanced threads busy), or \texttt{merge-and-backfill} (merge large task
counts from auxiliary queue).

\section{Further trace data}

\begin{figure}
 \begin{center}
  \includegraphics[width=0.8\textwidth]{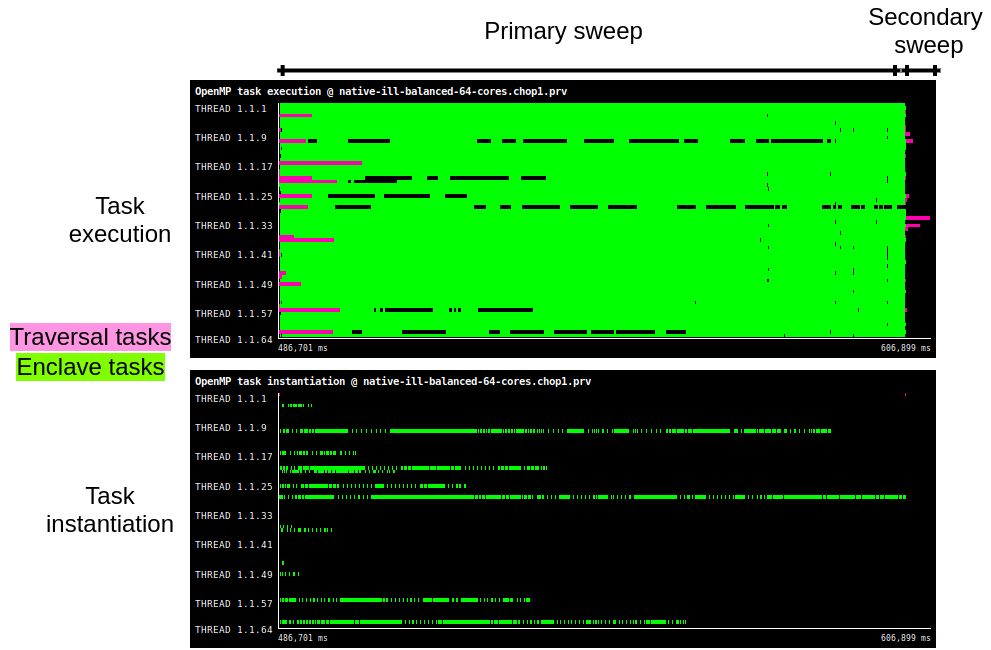}
 \end{center}
 \caption{
  Paraver trace showing task execution (top) and task spawning (bottom) pattern for an ill-balanced
  partitioning on Hawk. Native OpenMP tasks are used for this enclave tasking. Intel's OpenMP (ICC 19.1.0) was used here.
  \label{figure:traces:Hawk} 
 }
\end{figure}

To support claims on disadvantageous task execution patterns for our algorithm,
we study the ill-balanced setup on Hawk with native OpenMP tasks and collect
Paraver traces (Figure \ref{figure:traces:Hawk}).
Few cores host long, BSP-type traversal tasks.
The hardcoded ill-balancing imposes this pattern.
These few BSP-type traversal tasks serve as task producers, i.e.~they are
responsible for the overwhelming majority of enclave task productions.

On Hawk, the enclave tasks are reasonably balanced, as the idle threads steal
them once they hit the \texttt{taskwait} of the primary sweep.
The trace furthermore clarifies that the task production is spread out over the
primary traversal, i.e.~once the task count exceeds a certain threshold, the
producer tasks immediately process their tasks.

Most of the enclave tasks are processed before we step over the
\texttt{taskwait} and thus do not overlap into the secondary traversal.
Only very few of them remain.
The secondary traversal thus becomes very ill-balanced.
\end{document}